\documentclass[12pt]{article}
\usepackage{graphicx,epsfig}
\usepackage{color}
\usepackage{cite}
\title{Enhancement of the $d$-wave pairing correlations by charge and spin
ordering in the spin-one-half Falicov-Kimball model with Hund and Hubbard~coupling}
\author{Pavol Farka\v sovsk\'y\\
Institute  of  Experimental  Physics,  Slovak   Academy   of
Sciences\\
Watsonova 47, 043 53 Ko\v {s}ice, Slovakia}
\date{}
\begin{document}
\baselineskip=24pt
\maketitle

\begin{abstract}
Projector Quantum-Monte-Carlo Method is used to examine effects 
of the spin-independent $U_{fd}$ as well as spin-dependent $J_z$ 
Coulomb interaction between the localized $f$ and itinerant $d$ electrons 
on the stability of various types of charge/spin ordering and superconducting 
correlations in the spin-one-half Falicov-Kimball model with Hund and Hubbard 
coupling. The model is studied for a wide range of $f$ and $d$-electron
concentrations and it is found that the interband interactions $U_{fd}$
and $J_z$ stabilize three basic types of charge/spin ordering, and namely,
(i) the axial striped phases, (ii) the regular $n$-molecular phases and 
(iii) the phase separated states. It is shown that the $d$-wave pairing correlations
are enhanced within the axial striped and phase separated 
states, but not in the regular phases. Moreover, it was found 
that the antiferromagnetic spin arrangement within the chains further enhances the 
$d$-wave paring correlations, while the ferromagnetic one has a fully opposite effect.             
\end{abstract}

\newpage
\section{Introduction}
In the past decades, a considerable amount of effort has been 
devoted to understand a formation of an inhomogeneous charge and spin
stripe order in strongly correlated electron systems as well as its
relation to superconductivity. The motivation was clearly due to the 
observation of  such an ordering in doped nickelate~\cite{Ni}, 
cuprate~\cite{Cu} and cobaltate~\cite{Co} materials, some of which 
constitute materials that exhibit high-temperature superconductivity. 
The Hubbard and $t-J$ models have been used the most frequently in 
the literature to study the problem of stripe formation~\cite{Oles1,White}. 
These studies showed on two possible explanations of formation 
of inhomogeneous spatial charge and spin ordering. According 
to the first explanation the stripe phases arise 
from a competition between the tendency to phase 
separate (the natural tendency of the system) and the long-range 
Coulomb interaction~\cite{Emery}. Contrary to this explanation, 
White and Scalapino proposed~\cite{White} a new mechanism that does 
not require the long-range interactions and according to which 
the stripe  order arises from a competition between 
kinetic and exchange energies.  
Much simpler mechanism of a charge stripe formation 
in strongly correlated systems has been found by Lemanski 
et al.~\cite{Lemanski1} within the spinless Falicov-Kimball 
model (FKM). This model~\cite{Falicov} combines only the kinetic energy 
of itinerant $d$ electrons with the on-site Coulomb interaction
between the itinerant $d$ and localized $f$ electrons
and thus various types of charge orderings determined within this 
model for different $f$ and $d$ electron dopings are simply a compromise
between these two tendencies. 

However, the spinless version of the FKM, although non-trivial, is not 
able to account for all aspects of real experiments. For example, 
many experiments show that a charge superstructure
is accompanied by a magnetic superstructure~\cite{Ni,Cu}.
In order to describe both types of ordering in the unified picture
Lemanski~\cite{Lemanski2} proposed a simple model based on a 
generalization of the spin-one-half FKM that besides the local 
Coulomb interaction between $f$ and $d$ electrons takes into account
also the anisotropic, spin-dependent local interaction (of the Ising type)
that couples the localized and itinerant subsystems. 
It was found that this model is able to describe various types of
charge and spin orderings observed experimentally in strongly
correlated systems, including the diagonal and axial charge stripes 
with antiferromagnetic or ferromagnetic arrangement of spins 
within the lines. 

In spite of this fact, the extended model is still
oversimplified to describe all details of real materials. The major
simplification consists of omitting the Hubbard interaction $U_{dd}$
between the itinerant electrons of opposite spins that is crucial
if one interests, for example, in the superconducting correlations 
in the system. Since the main goal of this paper is to examine 
the impact of charge and spin superstructures on the superconducting
correlations we have extended the Lemanski's model by the Hubbard 
interaction term. From this point of view, the model considered 
here, for a description of superconducting correlations in the strongly 
correlated system, is the spin-one-half FKM
extended by the anisotropic spin-dependent interband interaction of the Ising 
type (between $f$ and $d$ electrons) and the intraband Coulomb interaction   
of the Hubbard type that acts between two $d$ ($f$) electrons of 
opposite spins.

It should be noted that despite an enormous 
research activity in the past the relation between the charge/spin 
ordering and the superconductivity is still controversial 
(an excellent review of relevant works dealing with this subject 
can be found in~\cite{Oles2}). A considerable progress in this field 
has been achieved recently by Maier et al.~\cite{Maier} and  Mondaini 
et al.~\cite{Mon}. Both groups studied the two-dimensional Hubbard model,
in which stripes are introduced externally by applying a spatially 
varying local potential $V_i$, and they found a significant enhancement 
of the d-wave pairing correlations. However, it should be noted
that the potential $V_i$ is phenomenological and as such has no direct 
microscopic origin that corresponds to a degree of freedom
in the actual materials. The advantage of our approach,  
based on the generalized spin-one-half FKM with $U_{fd}$, $J_z$ and $U_{dd}$ interactions is that  
the charge/spin stripes are present in the model intrinsically,
even in its $U_{dd}=0$ limit.

\section{Model}
The Hamiltonian of the model considered in this paper has the form  
\begin{eqnarray}
H&=&-t\sum_{\langle i,j\rangle\sigma}d^+_{i\sigma}d_{j\sigma} 
+ U_{fd}\sum_{i\sigma\sigma'}f^+_{i\sigma}f_{i\sigma}d^+_{i\sigma'}d_{i\sigma'}
+ J_z\sum_{i\sigma}(f^+_{i-\sigma}f_{i-\sigma} - 
f^+_{i\sigma}f_{i\sigma})d^+_{i\sigma}d_{i\sigma} 
\nonumber\\
&+& U_{dd}\sum_i d^+_{i\uparrow}d_{i\uparrow}d^+_{i\downarrow}d_{i\downarrow} \ ,
\end{eqnarray}
where $f^+_{i\sigma}, f_{i\sigma}$ are the creation and annihilation 
operators for an electron of spin $\sigma=\uparrow, \downarrow$ in the 
localized state at lattice site $i$ and $d^+_{i\sigma}, d_{i\sigma}$ are 
the creation and annihilation operators of the itinerant electrons in 
the $d$-band Wannier state at site~$i$.

The first term of (1) is the kinetic energy corresponding to quantum-mechanical 
hopping of the itinerant $d$ electrons between sites $i$ and $j$. These intersite
hopping transitions are described by the matrix elements $t_{ij}$, which are 
$-t$ if $i$ and $j$ are the nearest neighbours and zero otherwise.
The second term represents the 
on-site Coulomb interaction between the $d$-band electrons with density 
$n_d=N_d/L=\frac{1}{L}\sum_{i\sigma}d^+_{i\sigma}d_{i\sigma}$ and the localized
$f$ electrons with density  
$n_f=N_f/L=\frac{1}{L}\sum_{i\sigma}f^+_{i\sigma}f_{i\sigma}$, where $L$ is the 
number of lattice sites. The third term is the above mentioned anisotropic, 
spin-dependent local interaction of the Ising type between the localized 
and itinerant electrons that reflects the Hund's rule force. And finally,
the last term is the ordinary Hubbard interaction term for the itinerant 
electrons from the $d$ band. Moreover, it is assumed that the on-site Coulomb 
interaction between $f$ electrons is infinite and  so the double occupancy 
of $f$ orbitals is forbidden.

This model has several different physical interpretations
that depends on its application. As was already mentioned above, it can be
considered as the spin-one-half FKM extended by the Hund 
and Hubbard interaction term. On the other hand, it can be also considered
as the Hubbard model in the external potential generated by the 
spin-independent Falicov-Kimball term and the anisotropic
spin-dependent Hund term. Very popular interpretation of the model
Hamiltonian (1) is its $(U_{dd}=0)$ version that has been introduced by 
Lemanski~\cite{Lemanski2} who considered it as the minimal model of charge and 
magnetic ordering in coupled electron and spin systems. Its attraction 
consists in this that without the Hubbard interaction term $(U_{dd}=0)$ 
the Hamiltonian (1) can be reduced to the single particle Hamiltonian.
Indeed, using the fact that the $f$-electron occupation number 
$f^+_{i\sigma}f_{i\sigma}$ of each site $i$  commutes  with the Hamiltonian
(1),  it can be replaced by the classical variable $w_{i\sigma}$ taking only 
two values: $w_{i\sigma}=1$ or 0, according
to whether or not the site $i$ is occupied by the localized $f$ electron
and so the Hamiltonian (1) in absence of $U_{dd}$ term can be written as 
\begin{equation}
H=\sum_{ij\sigma}h^{(\nu)}_{ij}d^+_{i\sigma}d_{j\sigma},
\end{equation}
where $h^{(\nu)}_{ij}=t_{ij}+(U_{fd}w_i+J_z\nu{s_i})\delta_{ij}$,
$w_i=w_{i\uparrow}+w_{i\downarrow}=0,1$,
$s_i=w_{i\uparrow}-w_{i\downarrow}=-1,1$ 
(we remember that the double occupancy of $f$ orbitals 
is forbidden) and $\nu=\pm 1$. 
Thus for a given $f$-electron  $w=\{w_1,w_2,\dots,w_L\}$
and spin configuration $s=\{s_1,s_2,\dots,s_L\}$
the investigation of the model (2) is reduced to the investigation of the 
spectrum of $h^{(\nu)}$ for different $f$ electron/spin distributions. 
This can be performed exactly, over the full set of $f$-electron/spin 
distributions or approximatively. One such approximative 
method has been introduced in our previous papers~\cite{Fark1,Cenci} and it was shown
that it is very effective in description of ground-state properties
of the model Hamiltonian (2) and its extensions. The method consists
of the following steps: 
i) Chose a trial configuration $c=\{w;s\}$.
(ii) Having $U_{fd}$, $J_z$ and the total number of electrons
$N=N_f+N_d$ fixed, find all eigenvalues $\lambda^{(\nu)}_k$ of $h^{(\nu)}$. 
(iii) For a given $N_f=\sum_iw_i$ determine the ground-state energy
$E(c)$ of a particular $f$-electron/spin configuration by filling 
in the lowest $N_d=L-N_f$ one-electron levels $\lambda^{(\nu)}_k$.
(iv) Generate a new configuration $c'$ by moving a randomly
chosen electron to a new position which is chosen also as random
(or equivalently by flipping the randomly chosen spin).
(v) Calculate the ground-state energy $E(c')$. If $E(c')<E(c)$   
the new configuration is accepted, otherwise $c'$ is rejected.   
Then the steps (ii)-(v) are repeated until the convergence
(for given $U_{fd}$ and $J_z$ ) is reached.
Of course, one can move (flip) instead of one electron in step (iv)
simultaneously two or more electrons (spins), thereby the convergence 
of method is improved. For the $U_{dd} > 0$ case studied in the current 
paper we use the quantum variant of this method, that differs
from the classical one only in the step (iii) where the ground state
energy of the full Hamiltonian (1) has to be calculated now by a
quantum method. Here we used the exact diagonalization Lanczos 
method~\cite{Lanczos} for clusters less than $L=16$ sites and the Projector 
Quantum Monte Carlo method~\cite{PQMC} for larger clusters. 

Having the actual charge and spin distributions that minimize the ground state energy 
of the model Hamiltonian (1) all ground state observables can 
be calculated immediately. In the current paper we focus
our attention to the problem of superconducting correlations
in the ground state. Especially we study the influence of 
the  spin-independent Coulomb interaction $U_{fd}$ (the charge ordering) 
and the anisotropic spin-dependent interaction $J_z$ (the spin ordering) on 
the superconducting correlation function with $d_{x^2-y^2}$ wave symmetry defined 
as~\cite{Fettes}

\begin{equation}
C_d(r)=\frac{1}{L}\sum_{i,\delta,\delta'}g_{\delta}g_{\delta'}\langle
d^+_{i\uparrow}d^+_{i+\delta\downarrow}d_{i+\delta'+r\downarrow}d_{i+r\uparrow}
\rangle,
\end{equation}
where the factors $g_{\delta},g_{\delta'}$ are 1 in x-direction and
-1 in y-direction and the sums with respect to $\delta,\delta'$ are
independent sums over the nearest neighbors of site $i$.

However, on small clusters the above defined correlation function 
is not  a good measure for superconducting correlations,
since contains also contributions from the one particle 
correlation functions 
\begin{equation}
C^{\sigma}_0(r)=\frac{1}{L}\sum_{i}\langle d^+_{i\sigma}d_{i+r\sigma}
\rangle,
\end{equation}
that yield nonzero contributions
to $C_d(r)$ even in the noninteracting case. 

For this reason we use as the true measure for superconductivity
the vertex correlation function
\begin{equation}
C^{v}_d(r)=C_d(r)-\sum_{\delta,\delta'}g_{\delta}g_{\delta'}
C^{\uparrow}_0(r)C^{\downarrow}_0(r+\delta-\delta')
\rangle,
\end{equation}

and its average

\begin{equation}
C^{v}_d=\frac{1}{L}\sum_{i}C^{v}_d(i).
\end{equation}

\section{Results and discussion}
We have started our study with the case $J_z=0$, that is slightly simpler
from the numerical point of view, since in this case it is necessary
to work only with the charge degrees of freedom. Unlike previous 
studies~\cite{Lemanski2,epjb,pssb,wrzodak}, that have been done exclusively for
$U_{dd}=0$, we consider here also the Hubbard interaction term with the intraband
Coulomb interacting $U_{dd}=2$.  To reveal effects  of the interband Coulomb 
interaction between localized and itinerant electrons $U_{fd}$ 
on the formation of the various types of charge ordering, we have 
analyzed model for a wide range of $U_{fd}$ values ($U_{fd}=0,0,5,1,2,3 \dots 8$)
at all possible combinations of half and quarter $f$ and $d$ electron
fillings on cluster of $L=8 \times 8$ sites. The results of our numerical 
calculations obtained within the method described in detail above 
are summarized in Fig.~1. There is displayed the complete list of $f$-electron
configurations that minimize ground state energy of the model for
different values of $U_{fd}, n_f$ and $n_d$.
\begin{figure}[h!]
\begin{center}
\includegraphics[width=16cm]{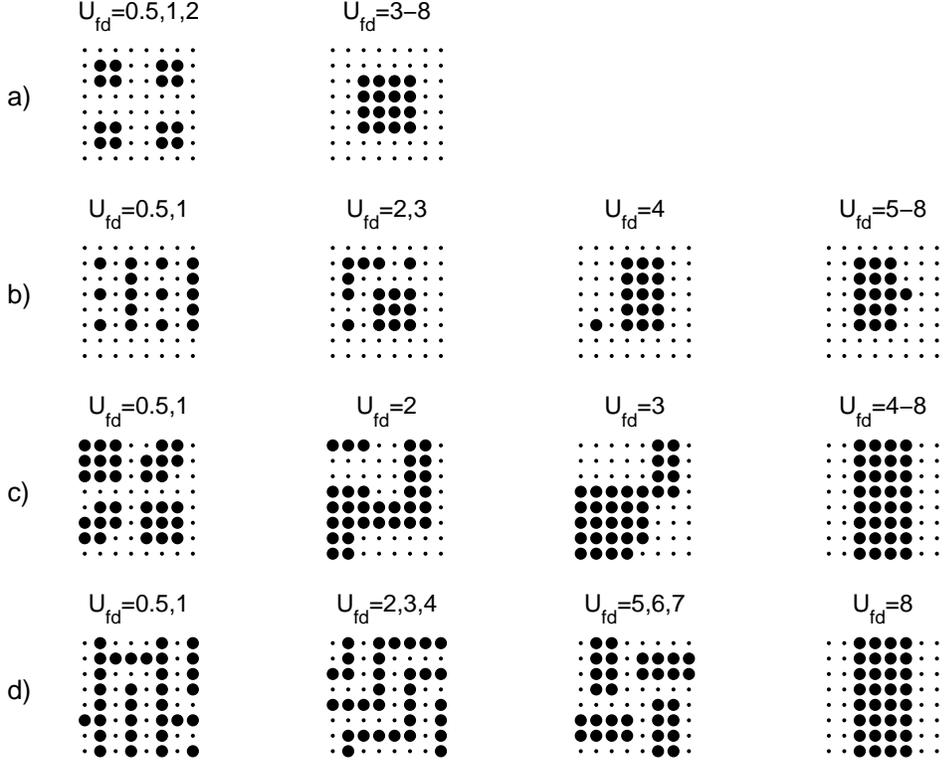}
\end{center}
\caption{\small Ground states of the model (1) calculated for $U_{dd}=2, J_z=0, 
L=8 \times 8$  at different values of the interband Coulomb interaction $U_{fd}$ and
different $f$ and $d$ electron filings: a) $n_f=1/4, n_d=1/4$;
b) $n_f=1/4, n_d=1/2$; c) $n_f=1/2, n_d=1/4$; d) $n_f=1/2, n_d=1/2$.   
Here the dark circles represent the $f$ electrons 
and dots the empty sites.}
\label{fig1}
\end{figure}
Analysing these results one can see some general trends
in formation of charge ordering. For sufficiently large $U_{fd}$
there is an obvious tendency to the phase separation for all examined 
$f$ and $d$ electron fillings, while in the opposite limit there is 
a tendency to form the axial stripes (for $n_f=1/4,n_d=1/2$ and 
$n_f=1/2,n_d=1/2$) or regularly distributed "n-molecules" of $f$-electrons
for ($n_f=1/4,n_d=1/4$ and $n_f=1/2,n_d=1/4$).
Between these three basic types of ground state configurations there
is a limited number of intermediate phases through which the low $U_{fd}$ 
phases transform to high $U_{fd}$ ones.

Of course, the charge ordering in the $f$-electron subsystem 
will have the strong impact on the $d$ electrons. To show what happens with 
the $d$ electrons we have calculated directly the $d$-electron on-site 
occupation $n_i=\sum_{\sigma}\langle d^+_{i\sigma}d_{i\sigma}\rangle$ for 
several representative $f$-electron distributions from Fig.~1, including the 
regular, phase separated and axial striped phases. 
\begin{figure}[h!]
\begin{center}
\includegraphics[width=12cm]{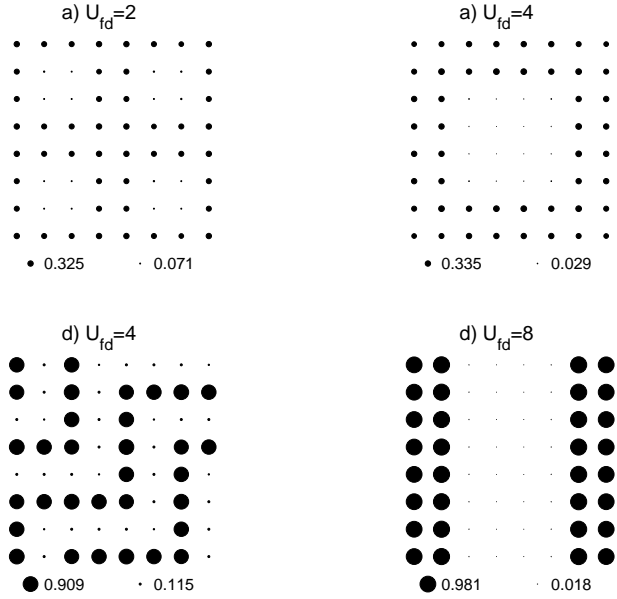}
\end{center}
\caption{\small The $d$-electron on-site occupation 
$n_i=\sum_{\sigma}\langle d^+_{i\sigma}d_{i\sigma}\rangle$
calculated for selected $f$-electron configurations from Fig.~1. The case a)
corresponds to $n_f=1/4, n_d=1/4$ and the case d) to $n_f=1/2, n_d=1/2$.
The on-site occupation $n_i$ is represented by filled-circles ($\bullet$), 
where the radius of circle on site $i$ is proportional to $n_i$.}
\label{fig2}
\end{figure}
It is seen (see Fig.~2) that 
$d$ and $f$ electrons exhibit the inverse site occupancy, which is obviously 
a consequence of the local Coulomb interaction $U_{fd}$ that prefers states 
without double $d$-$f$ site occupancy. As a result, the majority of $d$ 
electrons reside on the empty sites and only a fraction of them 
(due to the quantum mechanical hopping) share the same sites with $f$
electrons. This leads not only to the inverse $f$ and $d$ site occupancy, 
but also to the inverse charge patterns in the $f$ and $d$ electron 
subsystems and thus the axial stripes (the phase separation) in the $f$ 
electron subsystem are accompanied by the axial stripes (the phase separation)
in the $d$-electron subsystem.

The average vertex correlation functions corresponding to the ground state 
configurations from Fig.~1 are displayed in Fig.~3. Let us now discuss 
different cases in detail. 
\begin{figure}[h!]
\begin{center}
\includegraphics[width=14cm]{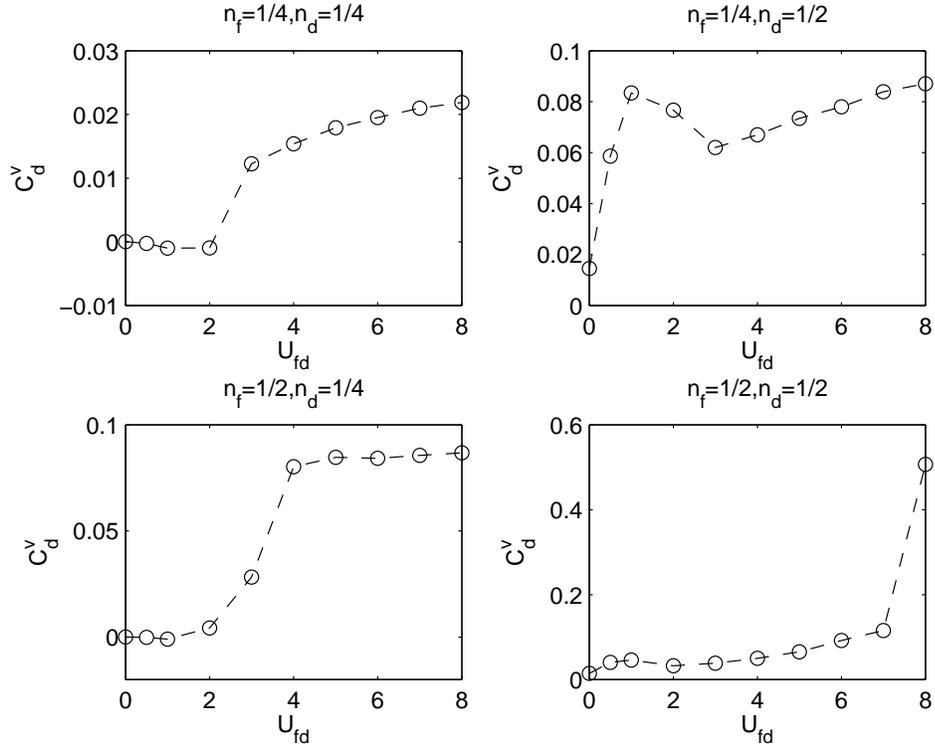}
\end{center}
\caption{\small Average vertex correlation function $C^v_d$ with $d_{x^2-y^2}$-wave
symmetry as a function of the interband Coulomb interaction $U_{fd}$ calculated 
for corresponding ground states from Fig.~1.}
\label{fig3}
\end{figure}
For $n_f=1/4,n_d=1/4$ the average vertex correlation 
function exhibits extremely different behaviour in regions of 
stability of the regular phase with the 4-molecules and the phase separated 
ground state. While in the first case, the correlation function is small and with
increasing $U_{fd}$ changes only slowly, in the second case it is
significantly enhanced at the phase boundary and further increases 
with increasing $U_{fd}$. A fully different
picture is observed for $n_f=1/4,n_d=1/2$. Here the average vertex correlation 
function sharply increases with increasing $U_{fd}$ within the axial striped
phase, then slowly decreases within the intermediate phases and it is again 
enhanced in the phase separated region. For $n_f=1/2,n_d=1/4$ the system
exhibits a similar behaviour as for $n_f=1/4,n_d=1/4$ with an exception
that the vertex correlations in the separated phase are now four
times larger than in the $n_f=1/4,n_d=1/4$ case. In the last case
$n_f=1/2,n_d=1/2$ the average vertex correlation function changes slowly for both 
the axial striped as well regular phase, but it is enhanced 
dramatically in the separated phase. This enhancement is by the factor 6,
in comparison to the $n_f=1/4,n_d=1/2$ and $ n_f=1/2,n_d=1/4$ phases
and even by the factor 25 for the $n_f=1/4,n_d=1/4$ phase, what 
emphasizes the role of $f$  ($d$) electron doping on the superconducting
correlations. 

Thus we can conclude that the separated and axial striped phases 
enhance the average vertex correlations in the spin-one-half FKM
with $U_{fd}, J_z$ and $U_{dd}$ couplings, while the regular phases have 
only a small impact on the $d_{x^2-y^2}$ superconducting correlations
in this model. Unfortunately, for half and quarter $f$ and $d$ electron fillings 
analysed above we have found no pure axial striped phases, as
observed in the experiments for some cuprate, nikelate and cobaltate 
systems and therefore we turn our attention to the case
$n_f+n_d=1$ and $n_f+n_d=2$, where such phases have been 
widely observed for both $J_z=0$ as well as $J_z>0$, 
even in the limit $U_{dd}=0$.~\cite{pssb}

The results of our nonzero $U_{dd}$ calculations are presented in Fig.~4
for $U_{dd}=2, U_{fd}=4$ and $J_z=0.5$. Comparing these results with ones 
obtained in our previous paper~\cite{pssb} for $U_{dd}=0$, and the same values
of  $U_{fd}$ and $J_z$, one can see that they are identical 
practically for all examined $f$ electron concentrations for both
$n=1$ and $n=2$. 
\begin{figure}[h!]
\begin{center}
\includegraphics[width=16cm]{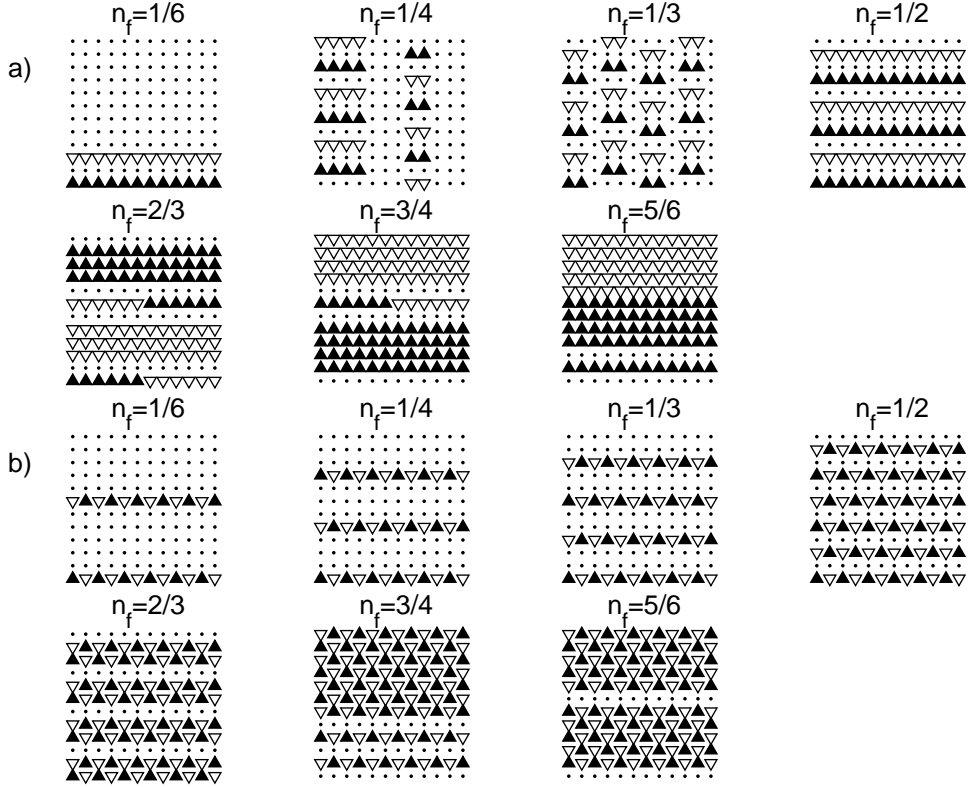}
\end{center}
\caption{\small Ground states of the model (1) calculated for $U_{dd}=2,
U_{fd}=4, J_z=0.5, L=12 \times 12$ and different values of the $f$-electron concentrations 
at (a) $n=n_f+n_d=1$ and (b) $n=n_f+n_d=2$. Here the spin 
up (down) of the $f$ electron is represented by a filled regular triangle 
(open inverted triangle).}
\label{fig4}
\end{figure}
The exceptions have been observed only for $n_f=1/4$ and $n_f=2/3$
($n=1$), where slightly different types of the ground states have been
identified for $U_{dd}=0$ and $U_{dd}>0$. 
This independently confirms the supposition made
in our previous papers~\cite{epjb,pssb}, based on the small cluster exact diagonalization
calculations, and namely that in the strong interaction $U_{fd}$ limit
($U_{fd} \geq 4$), the ground states of the model found for $U_{dd}=0$
persist as ground states also for nonzero $U_{dd}$, up to relatively 
large values ($U_{dd}\sim3$).   

Let us now summarize our numerical results for the ground states.  
(i) In all examined cases the ground states of the model (1) are 
non-polarized ($n_{f\uparrow}=n_{f\downarrow}$, $n_{d\uparrow}=n_{d\downarrow}$)
for both the $n_f+n_d=1$ as well as $n_f+n_d=2$ line.
(ii) For $n_f+n_d=1$ the ground states are either the regular distributions
of $f$ electrons ($n_f=1/3$) or the axial striped configurations, 
some of which can be phase separated ($n_f=1/6$ and $n_f=5/6$).
(iii) For $n_f+n_d=2$ only the axial striped configurations are the ground 
states of the model, and the phase separation is observed only for $n_f=3/4$.

The average vertex correlation functions calculated for a complete list
of ground state configurations from Fig.~4 are shown in Fig.~5 ($n=1$) 
and Fig.~6 ($n=2$). To see the enhancement of superconducting correlations
due to the interband Coulomb $U_{fd}$ and spin $J_z$ interaction 
against the ordinary Hubbard model with $U_{dd}=2$ we have plotted 
$C^v_d$ for both $U_{fd}=4, J_z=0.5$ and $U_{fd}=0, J_z=0$.
\begin{figure}[h!]
\begin{center}
\vspace{-0.5cm}
\includegraphics[width=14cm]{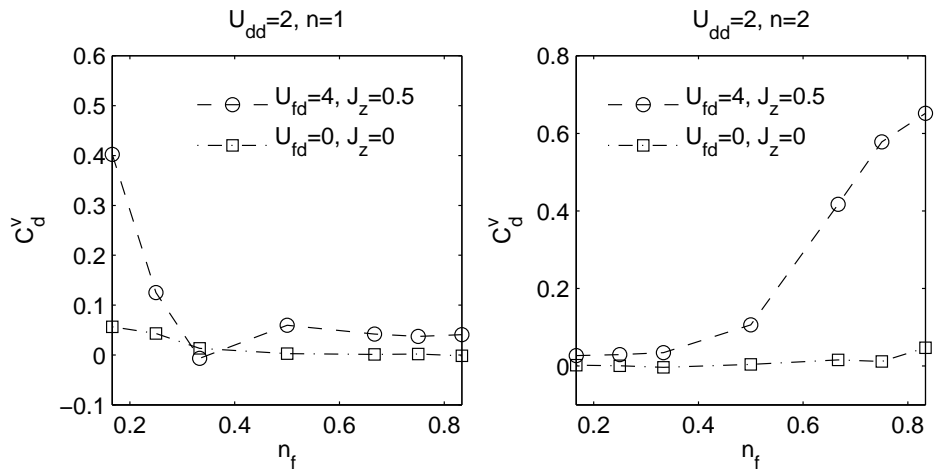}
\end{center}
\vspace{-2.5cm}
\caption{\small Average vertex correlation function $C^v_d$ as a function of $n_f$ calculated 
for corresponding ground states from Fig.~4 at $U_{fd}=4, J_z=0.5$ (the
dashed line) and $U_{fd}=0, J_z=0$ (the dash-dotted line).}
\label{fig5}
\end{figure}
One can see that the superconducting correlations for $n=1$ are considerably
enhanced for $n_f=1/6,1/4,1/2,2/3,3/4$ and $5/6$, while they are practically
unchanged for $n_f=1/3$. The ground state configurations
for $n_f=1/6,1/4,1/2,3/4$ and $5/6$ are either phase separated or 
axially striped and thus the enhancement of superconducting correlations 
in these phases is in accordance with our results discussed above for 
half and quarter $f$ and $d$ electron fillings. In accordance with 
these conclusions, it is also the result obtained for $n_f=1/3$. In 
this case the pairs of $f$-electrons are distributed regularly over the whole
two-dimensional lattice and thus the suppression of superconducting 
correlations is expected.

A fully different picture is observed 
for $n=2$, where the superconducting correlations are enhanced for 
all examined values of $f$-electron concentrations and they sharply
increase with $f$-electron doping (for example, for $n_f=5/6$ they are
enhanced by a factor 20 in comparison to the $U_{fd}=0$ and $J_z=0$ 
case). For all examined $n_f$ the ground states are the axially striped 
phases and thus these results are also consistent with our above mentioned 
conclusions.

To separate contributions to $C^v_d$ from $U_{fd}$ and $J_z$ we have plotted
in Fig.~6 the average vertex correlation function $C^v_d$ as a function of 
$n_f$ for $J_z=0.5$ as well as $J_z=0$.
\begin{figure}[h!]
\begin{center}
\includegraphics[width=14cm]{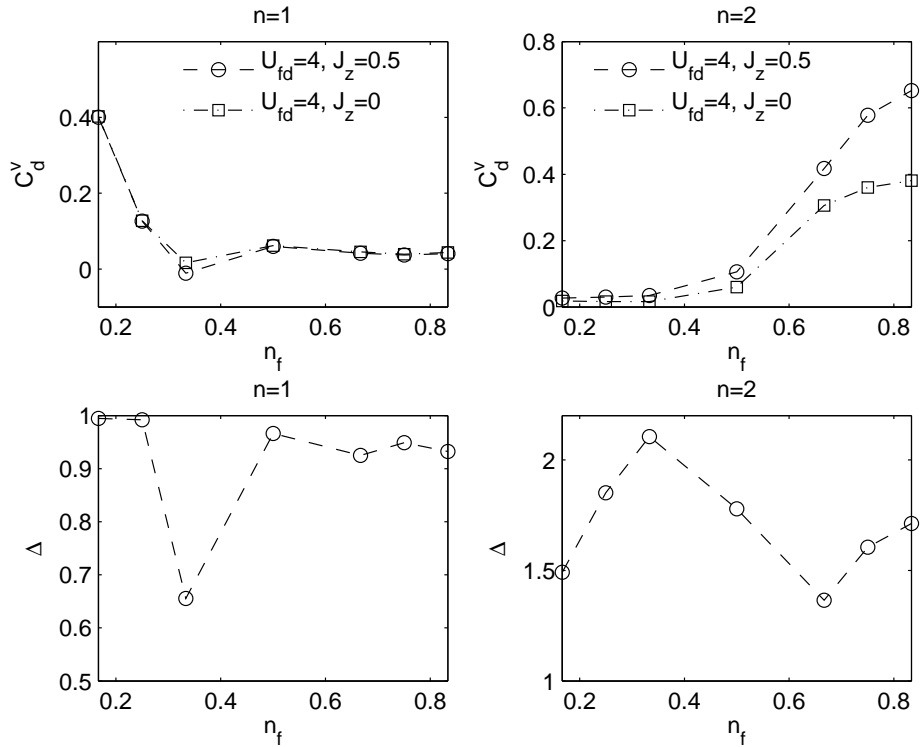}
\end{center}
\caption{\small Average vertex correlation function $C^v_d$ as a function of $n_f$ calculated 
for corresponding ground states from Fig.~4 at $U_{fd}=4, J_z=0.5$ (the
dashed line) and $U_{fd}=4, J_z=0$ (the dash-dotted line) for both $n=1$ and
$n=2$. The enhancement $\Delta$ corresponds to the ratio of the average vertex
correlation functions with and without the Ising coupling $J_z$, 
$\Delta=C^v_d(U_{fd}=4,J_z=0.5)/C^v_d(U_{fd}=4,J_z=0)$.}
\label{fig6}
\end{figure}
It is seen that the Hund constant suppresses the superconducting correlations
for $n=1$, the most obviously for $n_f=1/3$, while 
in the case $n=2$ the nonzero $J_z$ enhances considerably superconducting
correlations. Comparing the types of spin arrangement within the axial
phases for $n=1$ and $n=2$, the reason for such a different behaviour 
seems to be obvious. While for $n=1$ the spins are arranged ferromagnetically
within the individual lines, for $n=2$ they are arranged
antiferromagnetically. This implies that the antiferromagnetic correlations 
within the axial stripes enhance the superconducting correlations in the 
$d_{x^2-y^2}$ channel, while the ferromagnetic ones has the opposite effect.

In summary, the Projector Quantum Monte Carlo Method is used 
to examine effects of the interband Coulomb interaction $U_{fd}$ 
and the anisotropic spin-dependent interband interaction $J_z$ 
on the stability of various types of charge/spin ordering
and  superconducting correlations  in the two-dimensional 
spin-one-half FKM with Hund and Hubbard coupling. 
It is  found that for half and quarter 
$f$ and $d$ electron fillings the intraband Coulomb interaction 
$U_{fd}$ stabilized the axial striped or regular $n$-molecular phases 
for small and intermediate values of $U_{fd}$ and the phase separated 
$f$-electron distributions for $U_{fd}$ large. The superconducting 
vertex correlation functions were strongly enhanced within the 
axial striped and phase separated states, while they were small
within the regular $n$-molecular phases. Along the lines
$n_f+n_d=1$ and $n_f+n_d=2$ only the axial striped or $n$-molecular
charge phase were identified. For $n=1$ the $d$-wave correlations 
were enhanced for axial striped phases (with $n_f=1/6,1/4,1/2,3/4,5/6$) 
and suppressed for the periodic $n_f=1/3$ phase. Unlike this case, 
for $n=2$ the $d$-wave vertex correlations were enhanced for all 
examined $f$-electron fillings and 
the most significantly for $n_f \to 1$. Moreover, it was found
that the superconducting correlations in the axial striped phases
were strongly influenced by spin arrangements within lines.
In particular, the antiferromagnetic spin arrangement (found 
for $n=2$) further enhances the d-wave paring correlations, 
while the ferromagnetic spin arrangement (found for $n=1$) 
has the opposite effect.

\vspace{0.4cm}
This work was supported by Slovak Research and Development Agency (APVV)
under Grant APVV-0097-12 and ERDF EU Grants under the contract No.
ITMS 26220120005 and ITMS26210120002.


\newpage

\begin{thebibliography}{999}
\bibitem{Ni} C. H. Chen, S.-W. Cheong and A. S. Cooper, Phys. Rev. Lett.
{\bf 71}, 2461 (1993); J. M. Tranquada, D. J. Buttrey, V. Sachan and J. E.
Lorenzo, Phys. Rev. Lett.~{\bf 73}, 1003 (1994); Phys. Rev. B~{\bf 52}, 
3581 (1995); V. Sachan {\it et al.}, {\it ibid.} {\bf 51}, 12742 (1995). 

\bibitem{Cu} J. M. Tranquada, B. J.
Sternlieb, J. D. Axe, Y. Nakamura and S. Uchida, Nature (London) {\bf 375}, 
561 (1995); Phys. Rev. B~{\bf 54}, 7489 (1996); Phys. Rev. Lett.~{\bf 78}, 
338 (1997); H. A. Mook, P. Dai and F. Dogan, Phys. Rev. Lett.~{\bf 88}, 
097004 (2002); 

\bibitem{Co} K. Takada, H. Sakurai, E. Takayama-Muromachi, F. Izumi, R. Dilanian, 
and T. Sasaki, Nature (London) {\bf 422}, 53  (2003)

\bibitem{Oles1}  A. M. Oles, Acta Phys. Pol. B {\bf 31}, 2963 (2000).   
\bibitem{White}  S. R. White and D. J. Scalapino, Phys. Rev. Lett. {\bf 80}, 1272 (1998).   
\bibitem{Emery}  V. J. Emery, S. A. Kivelson, and H. Q. Lin, Phys. Rev. Lett.
{\bf 64}, 475 (1990).

\bibitem{Lemanski1} R. Lemanski, J. K. Freericks and G. Banach, Phys. Rev. Lett.
{\bf 89}, 196403 (2002);

\bibitem{Falicov} L.M. Falicov and J.C. Kimball, Phys. Rev. Lett.
{\bf 22}, 997 (1969).

\bibitem{Lemanski2} R. Lemanski, Phys. Rev. B {\bf 71}, 035107 (2005).

\bibitem{Oles2} A.M. Oles, Acta Physica Polonica B {\bf 121}, 752 (2012). 

\bibitem{Maier}  T.A. Maier, G. Alvarez, M. Summers and T.C. Schulthess,  
Phys. Rev. Lett.  {\bf 104}, 247001 (2010).

\bibitem{Mon} R. Mondaini, T. Ying, T. Paiva and R.T. Scalettar, 
Phys. Rev. B {\bf 86}, 184506 (2012). 


\bibitem{Fark1} P. Farka\v{s}ovsk\'y, Eur. Phys. J.  B {\bf 20}, 209 (2001).

\bibitem{Cenci} H. \v{C}en\v{c}arikov\'a and P. Farka\v{s}ovsk\'y, 
Int. J. Mod. Phys. B{\bf 18}, 357 (2004).

\bibitem{Lanczos} E. Dagotto, Rev. Mod. Phys. {\bf 66}, 763 (1994).

\bibitem{PQMC} M. Imada and Y. Hatsugai, J. Phys. Soc. Jpn. 58, 3752 (1989).

\bibitem{Fettes} M. Fettes, I. Morgenstern and T. Husslein, Computer Physics
Communications {\bf 106}, 1 (1997).

\bibitem{epjb} P. Farka\v{s}ovsk\'y and H. \v{C}en\v{c}arikov\'a, 
Eur. Phys. J. B {\bf 47}, 517 (2005).

\bibitem{pssb} H. \v{C}en\v{c}arikov\'a and P. Farka\v{s}ovsk\'y, 
phys. stat. sol. (b) {\bf 245}, 2593 (2008).

\bibitem{wrzodak} R. Lemanski and J. Wrzodak, Phys. Rev. B {\bf 78}, 085118
(2008).



\end{thebibliography}
\end{document}